\documentclass[iop]{emulateapj}
\shorttitle{Shock wave structure}
\shortauthors{Tolstov et al.}        
\usepackage{enumerate}
\usepackage{amsmath}
\usepackage{color}

\def\e#1{$\cdot 10^{#1}$ }

\begin{document}

\title{Shock wave structure in astrophysical flows \\ with an account of photon transfer
}

\author{Alexey Tolstov\altaffilmark{1,2,*}, Sergei Blinnikov\altaffilmark{1,2,3}, Shigehiro Nagataki\altaffilmark{4}, Ken'ichi Nomoto\altaffilmark{1,5}}
\affil{\altaffilmark{1} Kavli Institute for the Physics and Mathematics of the Universe (WPI), The
University of Tokyo Institutes for Advanced Study, The University of Tokyo, 5-1-5 Kashiwanoha, Kashiwa, Chiba 277-8583, Japan}  
\affil{\altaffilmark{2} ITEP, 117218 Moscow, Russia} 
\affil{\altaffilmark{3} VNIIA, 127055 Moscow, Russia}
\affil{\altaffilmark{4} Astrophysical Big Bang Laboratory, RIKEN, Saitama 351-0198, Japan}

\email{$^{*}$ E-mail: alexey.tolstov@ipmu.jp}

%\submitted{Submitted to ApJ on December 03, 2014} % --- Accepted on  February xx, xxxx.} 
\submitted{Accepted for publication in Astrophysical Journal, July 23, 2015}

\begin{abstract}
\noindent 

For an accurate treatment of the shock wave propagation in high-energy astrophysical phenomena, such as
supernova shock breakouts, gamma-ray bursts and accretion disks, knowledge of radiative transfer plays a crucial
role. In this paper we consider one-dimensional (1D) special relativistic radiation hydrodynamics by solving the
Boltzmann equation for radiative transfer. The structure of a radiative shock is calculated for a number of shock
tube problems, including strong shock waves, and relativistic- and radiation-dominated cases. Calculations are
performed using an iterative technique that consistently solves the equations of relativistic hydrodynamics and
relativistic comoving radiative transfer. A comparison of radiative transfer solutions with the Eddington
approximation and the M1 closure is made. A qualitative analysis of moment equations for radiation is performed
and the conditions for the existence of jump discontinuity for non-relativistic cases are investigated numerically.

\end{abstract}

%     supernovae, shocks, X-ray transients

% }
%\vskip 0.6cm
%\vfill
%\noindent\rule{5cm}{1pt}\
%{$^*$ E-mail $<$alexey.tolstov@riken.jp$>$}

%\noindent\rule{5cm}{1pt}\

%***************************************************************
%***************************************************************
\section{Introduction}
\noindent

\footnotetext[5]{Hamamatsu Professor.}

There are a number of topical high-energy astrophysical
phenomena in which radiative transfer occurs in moving media:
supernovae, gamma-ray bursts (GRBs), jets from active
galactic nuclei, and collapsars. In all of these phenomena,
radiative transfer can play a crucial role in the dynamics and
should be included in simulations leading to the identification
of observation signatures.

One of the most striking examples of phenomena of this type
is supernova shock breakout. It produces a bright flash, caused
by a shock wave emerging on the surface of the star after the
phase of collapse or thermonuclear explosion in the interior.
Recent detections of supernova shock breakouts \citep{Schawinski2008, Gezari2008, Soderberg2008} require
more accurate theoretical models, which are usually constructed
numerically in non-relativistic treatment 
(see e.g. \citet{KleinChevalier1978, EnsmanBurrows1992, KellyKorevaar1995, Blinnikov1998, Blinnikov2000, Katz2010, Tominaga2011, Katz2012, Sapir2013, Sapir2014}).
Our previous study of supernova shock
breakout phenomena \citep{Tolstov2013} revealed the
importance of accurate light curves of Type Ibc supernovae,
where the velocity of matter at the epoch of shock breakout
becomes mildly relativistic. The shock wave can reach highly
relativistic velocities in exploding white dwarfs and hypernovae.
Better descriptions of supernova shock breakout phenomena
and the connection to supernovae with GRBs require a self-consistent
solution of relativistic radiation hydrodynamics and
detailed study of shock wave structure.
% Also the detailed study of a shock wave structure is important for better description of gamma-ray bursts and their afterglows.

In this paper we consider the problem of the structure of shock
waves that are coupled with radiation at an arbitrary velocity of
matter for a one-dimensional plane stationary shock. There are
many papers related to the shock wave structure problem 
\citep{Zeldovich1957, Raizer1957, 
Belokogne1959, ImshennikMorozov1964, Morozov1971, Belokogne1972, Weaver1976, Chapline1984, 
Bouquet2000, Drake2007, Coulombel2009, Kraiko2011, Vaytet2013}\footnotetext[6]{Note that the name (Belokogne) is often spelled 
(Belokon').}, 
but in all of them the
non-relativistic motion of the fluid ($v/c \ll 1$) or approximations
of radiation field are considered. A number of papers have
been published where the shock wave structure is calculated
numerically for relativistic velocities of matter
\citep{Farris2008, Zanotti2011, 
Fragile2012, Sadowski2013, Takahashi2013, McKinney2014}, 
but
they are based on the solution of radiation moments equations
using a closure condition. In some cases this approach provides
a relevant solution, but sometimes it is difficult to estimate the
relevance of the closure condition, so the exact solution of
kinetic equation is required. To resolve this uncertainty and to
estimate how accurate the previous studies are, we take into
consideration the radiative transfer equation. This approach
eliminates the closure condition and helps to clarify the role of
radiation in the shock wave structure at various flow velocities.

As a striking example of the influence of radiative transfer on
the structure of a radiative shock wave we will consider the
phenomenon of the disappearance of shock wave jump
discontinuity (viscous jump) in the radiation-dominated flow
for non-relativistic velocities. The disappearance of the shock
wave jump discontinuity occurs when the downstream ratio of
the radiation pressure $p_r$ to the pressure of the gas $p_g$ rises above a critical value $(p_r/p_g)_{cr}$. The previous theoretical estimations provide values of the critical downstream ratio from  $(p_r/p_g)_{cr} \simeq 4.45$ \citep{Belokogne1959, Weaver1976} to $(p_r/p_g)_{cr} \simeq 8.5$ in more accurate considerations of radiation moment equations in the Eddington approximation \citep{ImshennikMorozov1964}. The critical ratio $(p_r/p_g)_{cr}$ seems to be very sensitive to the closure condition and our numerical simulations and analytic estimations provide its value more accurately.

\section{Statement of the problem}
\noindent

The problem of radiative transfer in moving media in a
general formulation leads to the solution of the system of the
relativistic radiation hydrodynamics equations coupled with the
relativistic radiative transfer equation. In this paper we consider
one-dimensional (1D), plane, and stationary shock waves. Due
to the complexity of the problem a number of simplifications
are introduced: Kirchhoff’s law of thermal radiation, the
graybody opacity law, zero value scattering opacity, and the
equation of state for an ideal gas.

Thus, the complete system of equations describing the shock
wave structure consists of the equation of state, the equations
of radiation hydrodynamics, and the equation of radiative
transfer:
\begin{align}
  np = \rho \epsilon,
\label{eos}
\end{align}
where $n$ is the polytropic index,  % $\Gamma=1+1/n$, 
$\epsilon$ - specific internal energy,
\begin{align}
 \gamma\beta\rho &= U_1, \\
 \gamma^2\beta^2[\rho+(n+1)p] +p + P &= U_2, \\
 \gamma^2\beta[\rho+(n+1)p] + F &= U_3, 
\label{hydro}
\end{align}
\begin{align}
  \gamma(\mu_0+&\beta)\frac{\partial I_0(\mu_0)}{\partial x}-
\gamma^3(1-\mu_0^2)(\mu_0+\beta)\frac{d\beta}{dx}
 \frac{\partial I_0(\mu_0)}{\partial \mu_0}+  \nonumber \\
  &+4\gamma^3\mu_0(\mu_0+\beta)\frac{d\beta}{dx}I_0(\mu_0)=\eta_0-\chi_0 I_0(\mu_0).
 \label{rt}
\end{align}

Here the density of the matter $\rho$ and the pressure $p$ in the
equations are measured in the fluid frame. The radiative
transfer equation is written in the comoving frame and the
dependence of the specific intensity $I_0$ from the space variable is suppressed in the notation.
The specific intensity $I_0$ and lab-frame radiation moments $E,F,P$ are related as:

\begin{align}
\{E_0,F_0,P_0\} &= \int_{-1}^{1} I_0(\mu_0)\mu^k d\mu_0, \;\; k=0...2, \\
E &= \gamma^2(E_0+2\beta F_0+\beta^2 P_0), \\
F &= \gamma^2((1+\beta^2)F_0+\beta(E_0+P_0)), \\
P &= \gamma^2(P_0+2\beta F_0+\beta^2 E_0).
 \label{moments_transform}
\end{align}

The equation of radiative transfer in the comoving frame (\ref{rt}) \citep{Mihalas1980} after integration over angle $\mu_0$ can be replaced by the radiation moments equations coupled with a closure condition (see Appendix \ref{appa} for details):

\begin{align}
  \frac{dF}{dx}&=\frac{\kappa U_1}{\beta}\Big( 
a_R T^4+ \nonumber \\
&+\frac{\beta F(2-f-\beta^2)-P(1-f\beta^2)}{f-\beta^2}\Big),
 \label{momentF}
\end{align}
\begin{align}
  \frac{dP}{dx}=\frac{\kappa U_1}{\beta}\Big( 
(a_R T^4+P)\beta-F
 \Big),
 \label{momentP} 
\end{align}
\begin{equation}
P_0 = fE_0
\label{closure}
\end{equation}

Equations (\ref{momentF}-\ref{closure}) are easier to solve in comparison to
the radiative transfer equation (\ref{rt}), but the choice of the Eddington factor $f$ in general case is not obvious. To compare the solution of the radiative transfer equation below we will consider a couple of 
approximations of the Eddington factor: 

\begin{enumerate}[a).]
\item {\it Eddington approximation} (optically thick medium)
\begin{equation}
f = \frac{1}{3}
\end{equation}
\item {\it M1 closure} \citep{Minerbo1978,Levermore1984,Dubroca1999} (combines approximations for optically thick and optically thin media)
\begin{equation}
f = \frac{1}{3}\Big[5-2\sqrt{4-3\Big(\frac{F_0}{P_0}\Big)^2}\Big]
\end{equation}
\end{enumerate}

\section{Numerical solution and tests}
\noindent

  To investigate how the radiative transfer affects the shock
wave structure we consider a number of shock tube tests that
are widely used in relativistic radiation hydrodynamics \citep{Farris2008, Zanotti2011, Fragile2012, Sadowski2013, Takahashi2013, McKinney2014}. These test data are summarized in Table \ref{params}.
  
  A number of closure conditions are used: the Eddington approximation, the M1 closure and the exact solution of radiative transfer problem.

  We use {\sc RADA} code \citep{TolstovBlinnikov2003, Tolstov2010} to solve the radiative transfer equation in the
comoving frame (\ref{rt}). The code is based on the method of characteristics and can be applied to the motion
  of the fluid with arbitrary velocity. The code solves the Boltzmann form of
the transport equation using the method of characteristics and
can be applied to the motion of the fluid with an arbitrary
velocity. The code is multigroup, but because we consider a
gray medium, no spectra details are calculated. Hydro
equations are solved semi-analytically (see Appendix \ref{appa} for
details). The numerical space grid used for radiative transfer is
independent of the hydro grid and is based on an adaptive
algorithm to improve the performance of the characteristics
method and resolve the discontinuities of radiation quantities.
The hydro equation and the radiative transfer equation are
solved by an iterative algorithm. The radiation moments
equations (\ref{momentF}-\ref{momentP}) are integrated by Dormand-Prince method with adaptive step.
    
    { In all the tests, following \citet{Farris2008}, we use
dimensionless quantities to make a comparison with already
published results.
     The upstream gas density $\rho_L$ is set to unity in all the tests. The choice of the other upstream quantities $v_L,p_L,E_L$ depends on the case to be considered. For a non-relativistic shock $v \ll 1$, and for a relativistic shock $v$ is close to unity. The values of pressure $p_L$ and radiation energy $E_L$ determine how strong the shock is and whether or not the shock
is radiation-dominant. In order to find all downstream physical
quantities and to construct a shock tube configuration the
nonlinear equation is solved (see Appendix \ref{appa}). }
    
     {   In this dimensionless approach the density in CGS units is related to dimensionless density as follows: $\rho_{CGS}= \rho \, [a_R (\frac{p_L}{\rho_L} 
    \frac{m}{k_B}c^2)^4 / (E_L c^2) ]$, where $a_R$ is the radiation constant, $m$ is the mean mass of baryons in the fluid, $k_B$
is Boltzmann's constant and $c$ is the speed of light.
         }

\begin{table*}
\begin{minipage}{146mm}
\caption{ Test Parameters for Left(L) and Right(R) States, $\rho_L$ = 1. %$v_L = 0.01$, $\rho_L = 1$,  $p_L = 3 \cdot 10^{-8}$.
\label{params}}
\begin{center}
\begin{tabular}{llllllllll}
 \\[-7mm]
\hline
\hline
 \\[-3mm]
Test & n & $v_L$ & $p_L$ & $E_L$ & $\rho_R$ & $v_R$ & $p_R$ & $E_R$ & $\kappa^a$ \\%& $(p_r/r_g)_L$ & $(p_r/r_g)_R$ & $(p_R/p_L)-1$ & $\kappa$ & $(E_R/E_L)$ \\
%     & $T_{\rm eff}$, K & $R_{\tau=2/3}$, cm & $\int_0^{2 {\rm d}}Ldt$, erg\\
\hline
 \\[-3mm]
f1 & 1.5 & 1.50\e{-2}& 3\e{-5}&1.0\e{-8}& 2.40& 6.25\e{-3}& 1.61\e{-4} & 2.5\e{-7}& 0.4\\
f2 & 1.5 & 2.43\e{-1}& 4\e{-3}&2.0\e{-5}& 3.11& 8.01\e{-2}& 4.51\e{-2} & 3.5\e{-3}& 0.2\\
f3 & 2 & 9.95\e{-1}& 60     &2.0      & 7.99& 0.78      & 2.34\e{3}  & 1.1\e{3} & 0.3\\
f4 & 1.5 & 5.68\e{-1}& 6\e{-3}&1.8\e{-1}& 3.65& 0.18      & 3.58\e{-2} & 1.3      & 0.08\\
\hline
% \\[-3mm]
%variant %& $t$, d & $L_{\rm bol}$, erg/s  & $T_{\rm c}$, K
%
%\hline
% \\[-3mm]
%14E0.7     & 4.71\e{5} & 3.32\e{12}& 1.07\e{47}\\
%14E1       & 5.28\e{5} & 3.32\e{12}& 1.40\e{47}\\
%14E1.3     & 5.73\e{5} & 3.32\e{12}& 1.77\e{47}\\
%14E1.34R   & 6.26\e{5} & 2.74\e{12}& 1.36\e{47}\\
%14E1.4R6   & 5.30\e{5} & 3.95\e{12}& 2.38\e{47}\\
\hline
\end{tabular}

\begin{tabular}{lllll}
\\[0.5mm]
\hline
\hline
 \\[-3mm]
Test &$(p_r/p_g)_L$ & $(p_r/p_g)_R$ & $(p_R/p_L)$ &  $(E_R/E_L)$  \\
%     & $T_{\rm eff}$, K & $R_{\tau=2/3}$, cm & $\int_0^{2 {\rm d}}Ldt$, erg\\
\hline
 \\[-3mm]
f1& 1.11\e{-4}& 5.19\e{-4}& 5.37 & 2.5\e{1} \\
f2& 1.67\e{-3}& 2.55\e{-2}& 11.2 &1.7\e{2}   \\
f3& 1.11\e{-2}& 1.61\e{-1}& 39.0 &5.6\e{2}   \\
f4& 10        & 1.20\e{1} & 5.98 &7.2       \\
\hline
% \\[-3mm]
%variant %& $t$, d & $L_{\rm bol}$, erg/s  & $T_{\rm c}$, K
%
%\hline
% \\[-3mm]
%14E0.7     & 4.71\e{5} & 3.32\e{12}& 1.07\e{47}\\
%14E1       & 5.28\e{5} & 3.32\e{12}& 1.40\e{47}\\
%14E1.3     & 5.73\e{5} & 3.32\e{12}& 1.77\e{47}\\
%14E1.34R   & 6.26\e{5} & 2.74\e{12}& 1.36\e{47}\\
%14E1.4R6   & 5.30\e{5} & 3.95\e{12}& 2.38\e{47}\\
\hline
\end{tabular}

\end{center}
%Note. The $T_{\rm eff}$ maximum almost coincides in time with the
%$L_{\rm bol}$ peak, while the $T_{\rm c}$ maximum occurs $\sim 100$ s earlier.
\end{minipage}
\end{table*}

  \begin{figure}
\includegraphics[width=84mm]{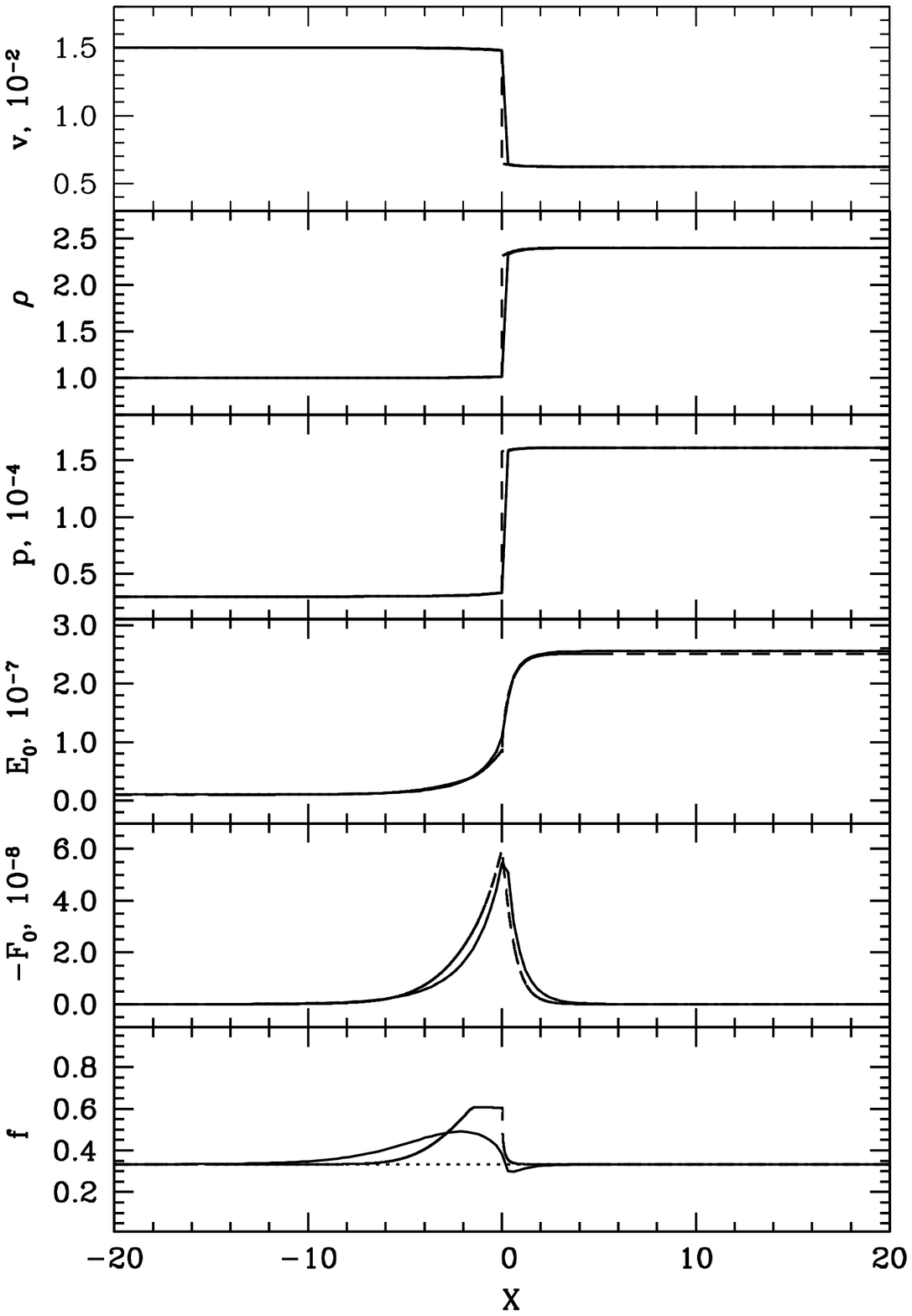}
\caption{ Profiles of $\beta$, $\rho$, $p$, $E_0$, $F_0$, $f$ for test f1 
at various radiation closure conditions. The dotted line is the Eddington approximation; the dashed line is the
M1 closure; and the solid line is the exact solution of the radiative transfer equation.
 }
\label{test1}
\end{figure}
  
    \begin{figure}
\includegraphics[width=84mm]{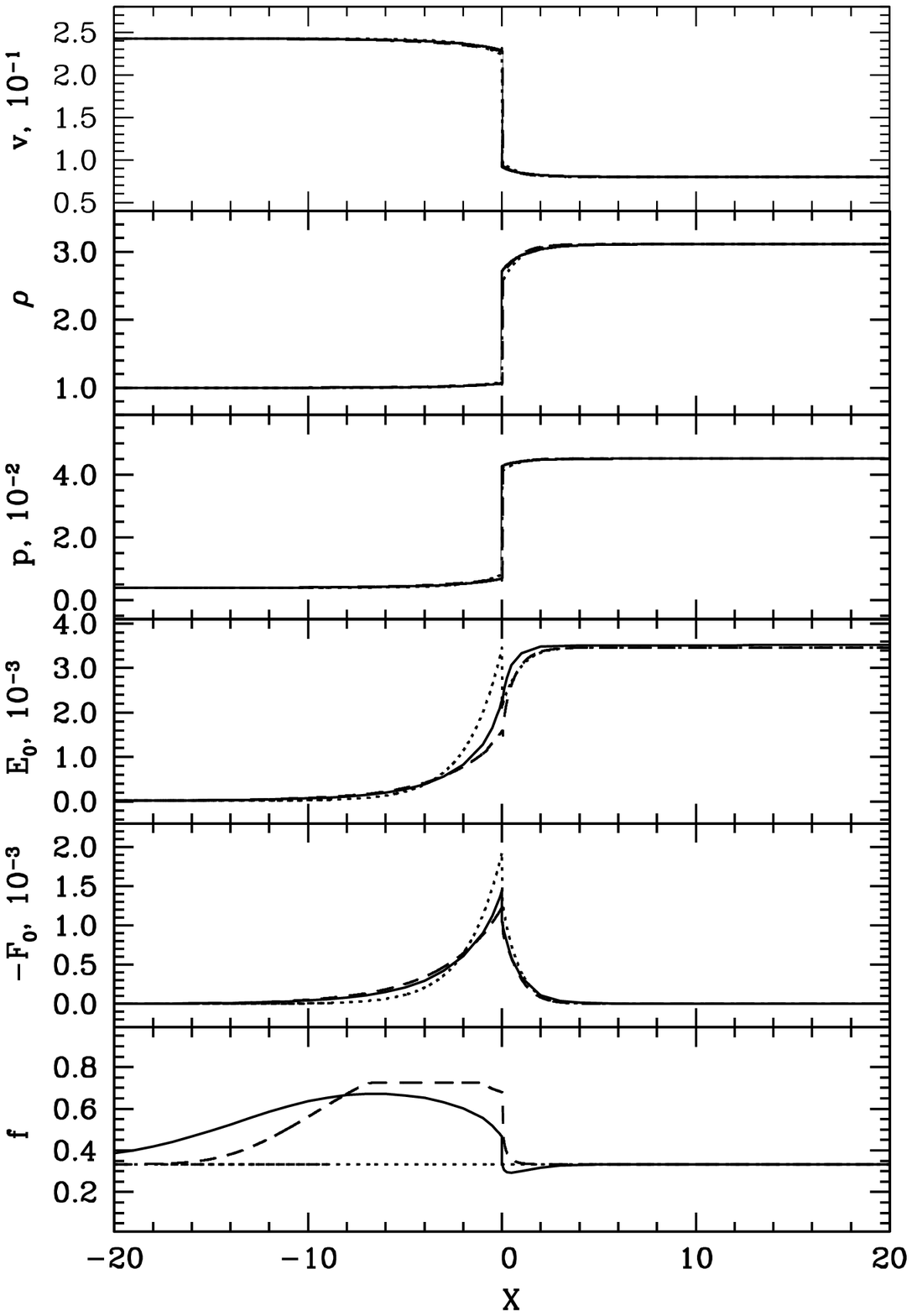}
\caption{ Profiles of $\beta$, $\rho$, $p$, $E_0$, $F_0$, $f$ for test f2 
at various radiation closure conditions. The dotted line is the Eddington approximation; the dashed line is
the M1 closure; and the solid line is the exact solution of the radiative transfer
equation.
 }
\label{test2}
\end{figure}

\begin{figure}
\includegraphics[width=84mm]{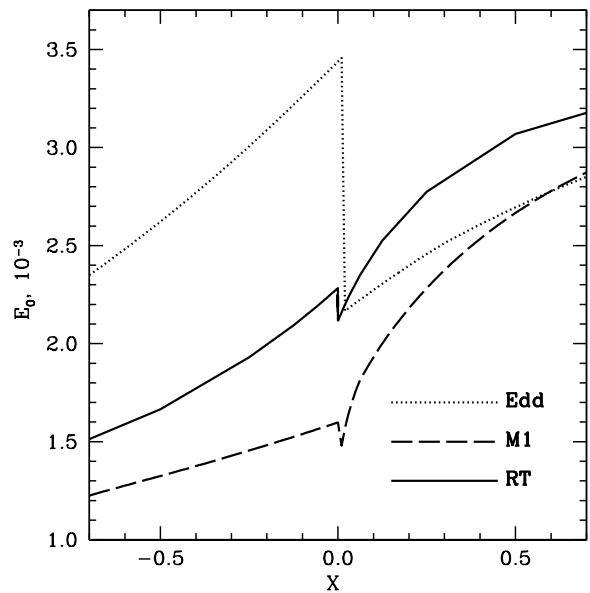}
\caption{ Radiation energy density ptofile near the discontinuity for test f2. The dotted line is the Eddington approximation; the dashed line is the
M1 closure; and the solid line is the exact solution of the radiative
transfer equation.
 }
\label{test2jmp}
\end{figure}

    \begin{figure}
\includegraphics[width=84mm]{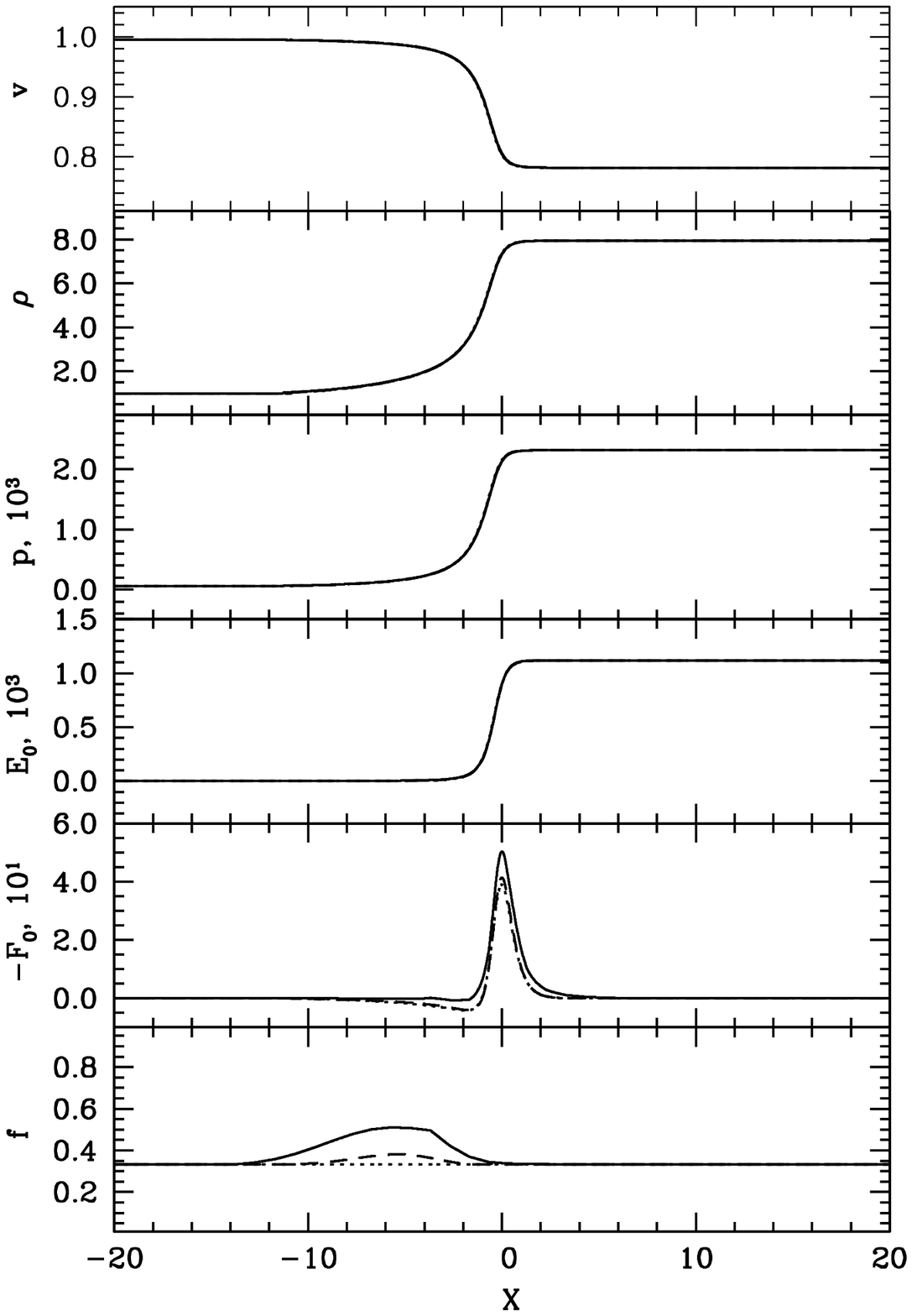}
\caption{ Profiles of $\beta$, $\rho$, $p$, $E_0$, $F_0$, $f$ for test f3 
at various radiation closure conditions. The dotted line is the Eddington approximation; the dashed line is the
M1 closure; and the solid line is the exact solution of the radiative
transfer equation.
 }
\label{test3}
\end{figure}    
  
    \begin{figure}
\includegraphics[width=84mm]{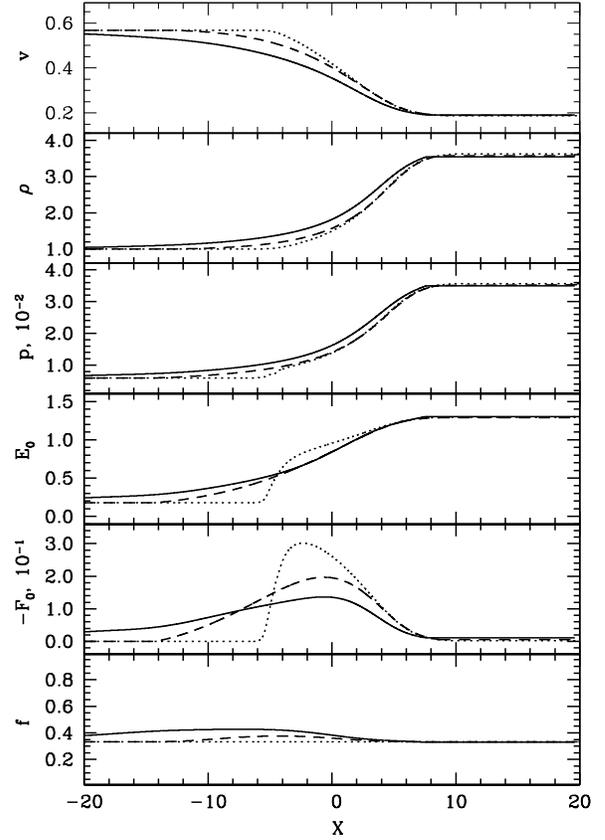}
\caption{ Profiles of $\beta$, $\rho$, $p$, $E_0$, $F_0$, $f$ for test f4 
at various radiation closure conditions. The dotted line is the Eddington approximation; the dashed line is the
M1 closure; and the solid line is the exact solution of the radiative
transfer equation. 
 }
\label{test4}
\end{figure}  

\subsection{Non-relativistic strong shock}
   In this test a strong, gas-pressure dominated, non-relativistic
shock is propagating into a cold gas. The exact solution of
radiative transfer and the M1 closure provide much higher
values of the Eddington factor f near the shock front than the
Eddington closure (Fig.\ref{test1}). Nevertheless this does not significantly change the hydro and radiation profiles. The high
value of the Eddington factor in the upstream has almost no
effect on the gas-pressure dominated, cold medium.

\subsection{Mildly-relativistic strong shock}
    {
The junction conditions for the radiation variables are the
continuity of the radiation flux $F$ and the radiation pressure $P$. But in the comoving frame these variables can become
discontinuous. This test reveals discontinuity not only in
hydrodynamic quantities but in the comoving radiation energy
density and the radiation energy flux (Fig. \ref{test2}). The jump of
the radiation energy density $E_0$ can roughly be estimated by a
simplified assumption of an optically thick medium and a
velocity profile as a step function at the discontinuity region.
Using an invariant frequency-integrated intensity $I/\nu^4$} for the radiation energy density jump $\Delta E_{0,d}$ we find:
\begin{align}
  \Delta E_{0,d} &= E_{L,d} - E_{R,d} {\approx} \nonumber \\
  &{\approx} \frac{1}{2}\Big[ E_R\int_{\beta_L}^{1}\frac{d\mu}{(1-\mu\beta)^4} 
  + E_L\int_{-\beta_L}^{1}d\mu \Big] -  \nonumber \\
  &- \frac{1}{2}\Big[ E_L\int_{-\beta_R}^{1}\frac{d\mu}{(1-\mu\beta)^4}
  + E_R\int_{\beta_R}^{1}d\mu \Big],  
\end{align}  
\begin{align}
 \beta = \frac{\beta_L-\beta_R}{1-\beta_L\beta_R}.
\end{align}  

  Substituting $\beta_L$ and $\beta_R$ from Table I one can estimate the upper limit of the jump as $\Delta E_{0,d} \simeq 3.5\cdot 10^{-4}$, which is in good agreement with our numerical calculation, yet several times lower than in the Eddington approximation {(Fig. \ref{test2jmp})}. The M1 closure gives a much better approximation of
relativistic transformations for this test.
  
Calculating this kind of discontinuity requires a fine space
grid for radiative transfer and we use an adaptive grid to
increase the number of points near the discontinuity region. We
note that the same discontinuity can be resolved for Test f1 as
well, but it is negligibly small due to non-relativistic velocities.

\subsection{Highly-relativistic wave}
This is a gas-dominated test, but no discontinuity is observed
(Fig. \ref{test3}) due to hot upstream gas. Similar to test f1, no
significant difference between the various closure conditions is
revealed in the profiles. The Eddington factor in the upstream
cannot affect radiation profiles because of high radiation energy
density in the downstream.

\subsection{Radiation-pressure dominated, mildly-relativistic wave}
This test is radiation-dominated, but the shock wave is not
strong and the radiation energy density in the downstream is
only several times higher than that in the upstream. This fact
leads to a larger dependence of radiation hydro profiles on
closure conditions (Fig. \ref{test4}). Both the hydro and radiation
profiles become smoother for the M1 closure in comparison
with the Eddington approximation, and are even smoother in
the exact calculation of the radiative transfer. Thus in this test
the solution is quite sensitive to the Eddington factor value.

\begin{table*}
\begin{minipage}{146mm}
\caption{Test Parameters at $v_L = 0.01$, $\rho_L = 1$,  $p_L = 3 \cdot 10^{-8}$, $n=1.5$.
\label{params2}}
\begin{center}
\begin{tabular}{lllllllllll}
\\[-7mm]
\hline
\hline
 \\[-3mm]
Test & $\lg{E_L}$ & $\rho_R$  & $v_R$,$10^{-3}$ & $p_R$,$10^{-5}$ & $E_R$ & {$(p_r/p_g)_L$} & $(p_r/p_g)_R$ & $(p_R/p_L)$ & $(E_R/E_L)$ \\
%     & $T_{\rm eff}$, K & $R_{\tau=2/3}$, cm & $\int_0^{2 {\rm d}}Ldt$, erg\\
\hline
 \\[-3mm]
s1  & -19 & 3.994 & 2.503 & 7.499 & 1.53\e{-8}& 1.11\e{-12} & 6.81\e{-5} & 2500& 1.53\e{11}\\
s2  & -18 & 3.996 & 2.502 & 7.496 & 1.52\e{-7}& 1.11\e{-11} & 6.79\e{-4} & 2498& 1.52\e{11}\\
s3  & -17 & 4.013 & 2.491 & 7.462 & 1.47\e{-6}& 1.11\e{-10} & 6.59\e{-3} & 2487& 1.47\e{11}\\
s4  & -16 & 4.143 & 2.413 & 7.212 & 1.13\e{-5}& 1.11\e{-9}  & 5.23\e{-2} & 2404& 1.13\e{11}\\
s5  & -15 & 4.569 & 2.188 & 6.314 & 4.50\e{-5}& 1.11\e{-8}  & 2.37\e{-1} & 2104& 4.50\e{10}\\
s6  & -14 & 5.183 & 1.929 & 4.869 & 9.61\e{-5}& 1.11\e{-7}  & 6.58\e{-1} & 1623& 9.61\e{9}\\
s7  & -13 & 5.764 & 1.734 & 3.383 & 1.46\e{-4}& 1.11\e{-6}  & 1.44       & 1127& 1.46\e{9}\\
s8  & -12 & 6.213 & 1.609 & 2.178 & 1.86\e{-4}& 1.11\e{-5}  & 2.85       & 726 & 1.86\e{8}\\
s9  & -11 & 6.519 & 1.533 & 1.330 & 2.14\e{-4}& 1.11\e{-4}  & 5.36       & 443 & 2.14\e{7}\\
s10 & -10 & 6.714 & 1.489 & 0.785 & 2.31\e{-4}& 1.11\e{-3}  & 9.83       & 261 & 2.31\e{6}\\
\hline
% \\[-3mm]
%variant %& $t$, d & $L_{\rm bol}$, erg/s  & $T_{\rm c}$, K
%
%\hline
% \\[-3mm]
%14E0.7     & 4.71\e{5} & 3.32\e{12}& 1.07\e{47}\\
%14E1       & 5.28\e{5} & 3.32\e{12}& 1.40\e{47}\\
%14E1.3     & 5.73\e{5} & 3.32\e{12}& 1.77\e{47}\\
%14E1.34R   & 6.26\e{5} & 2.74\e{12}& 1.36\e{47}\\
%14E1.4R6   & 5.30\e{5} & 3.95\e{12}& 2.38\e{47}\\
\hline
\end{tabular}
\end{center}
%Note. The $T_{\rm eff}$ maximum almost coincides in time with the
%$L_{\rm bol}$ peak, while the $T_{\rm c}$ maximum occurs $\sim 100$ s earlier.
\end{minipage}
\end{table*}

\section{Qualitative analysis}
\noindent

One can see that in some tests from the previous section we
find discontinuity, but in others all of the profiles are smooth.
The critical downstream radiation-to-gas pressure ratio $(p_r/p_g)_{cr}$,  at which the discontinuity disappears in nonrelativistic
consideration, has been investigated previously. In
consideration of shock jump conditions the ratio $(p_r/p_g)_{cr} \simeq 4.45$ for monoatomic gas (\citet{Belokogne1959,Weaver1976}; see also the detailed study in \citet{Belokogne1972} and the more correct original paper \citep{Belokogne1972r}). Using a more sophisticated approach for the Eddington approximation ($(p_r/p_g)_{cr} \simeq 8.5$, \citep{ImshennikMorozov1964}). In more a general formulation
$p_r/p_g=2+(3+2n)^{1/2}$ (\citet{Belokogne1972}, see also \citet{WeaverChapline1974} for a relativistic case), but in this paper we limit our consideration by the polytropic index $n=1.5$.

Using the solution of the radiative transfer problem the exact
value of this ratio can be calculated using our code.

In all of the calculations of the critical ratio the upstream gas
is supposed to be cold, its pressure is supposed to be low and
the fluid velocities should be non-relativistic

Our analysis is based on the approach described in \citet{ImshennikMorozov1964} paper. . In that paper, the approximate
solution of the problem and a qualitative analysis of equations (\ref{momentF}-\ref{momentP}) in the phase plane are performed in the
Eddington approximation. This approach gives the critical ratio $(p_r/p_g)_{cr} \simeq 8.5$ \citep{ImshennikMorozov1964}.

In this paper we extend the phase plane analysis to arbitrary
velocities (see Appendex \ref{appa}), take radiative transfer into
consideration, and construct a number of tests to analyze the
shock wave structure for low-pressure, cold upstream gas, and
specify the value of the critical ratio in the exact solution of the
radiative transfer equation.

The tests we have constructed (Table II) provide a set of
shock tube configurations from classical shock, with discontinuity
between the radiation-dominated shock waves up to
the downstream radiation-to-gas pressure ratio $p_r/p_g \simeq 10$. Upstream density, velocity, and pressure are the same for all of
the tests, radiation energy density is a variable parameter, and upstream pressure and temperature are negligible in comparison
with downstream pressure and temperature.

\begin{figure}
\includegraphics[width=84mm]{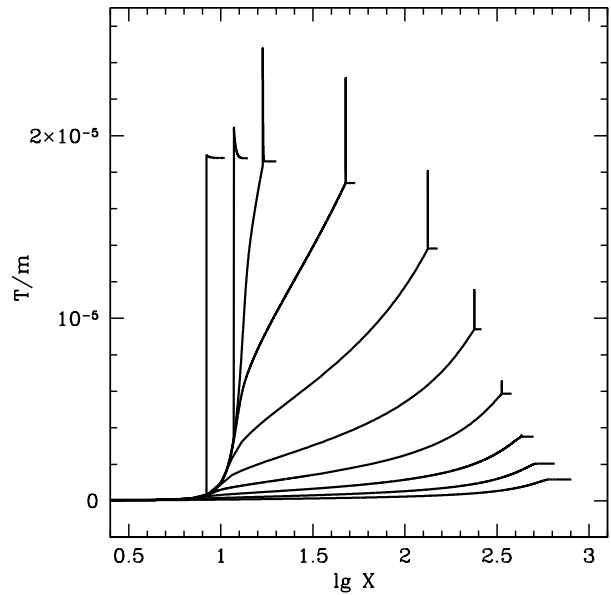}
\caption{Temperature profile of the shock front in the presence of radiation in
the Eddington approximation closure. The radiation influence $p_r/p_g$ increases
from the left to the right profile (Table I).
 }
\label{swedd2}
\end{figure}

\begin{figure}
\includegraphics[width=84mm]{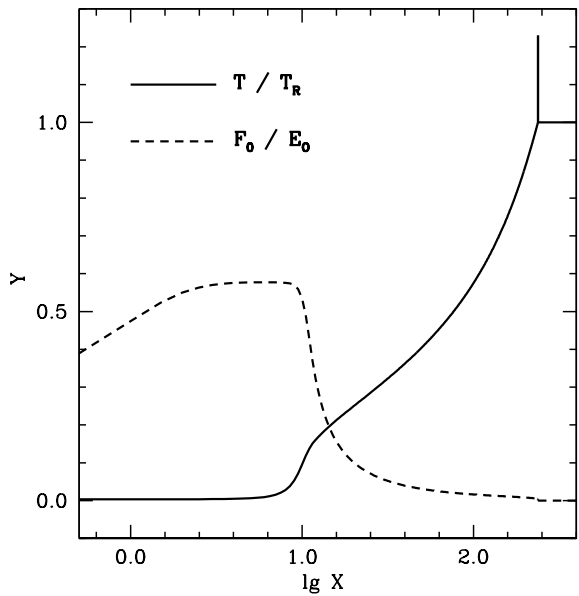}
\caption{Normalized temperature and radiation flux for test s4 (Table 1) in
solution with the M1 closure.
 }
\label{cmptfe}
\end{figure}

Our modeling of the shock front using the Eddington closure
reproduces Zeldovich spike behavior and the results of \citet{ImshennikMorozov1964} paper: i.e., 
the disappearance of the shock jump at $p_r/p_g \simeq 8.5$ (Fig. \ref{swedd2}). 
At $p_r/p_g>5$ the peak
temperature differs from the downstream temperature by a
fraction of a percent and there is not much difference in the
shock wave structure between the solutions with different
closure conditions.

{
We also performed the calculation of the profiles with the
M1 closure, but the result for the strong shock is almost the
same as for the Eddington closure (similar to tests f1-f3 in the
previous section) because radiation flux $F_0$ is comparable with
radiation density $E_0$ only in the part of the shock front where
the temperature is relatively low (see Figure \ref{cmptfe}).

Calculations with the radiation transfer equation do not
significantly change the profiles for the same reason: a high
Eddington factor affects only the low-energy part of the
temperature profiles (see Figures \ref{t1cmp}-\ref{t2cmp}). 

The critical ratio in the Eddington approximation $(p_r/p_g)_{cr} \simeq 8.5$ does not correspond 
to the value $(p_r/p_g)_{cr} \simeq 4.45$ found by \citet{Belokogne1959}. 
The approach described by \citet{ImshennikMorozov1964} seems to be more accurate because the
complete system of radiation hydrodynamics equations is
considered and as shown in their paper, the Eddington
approximation is more reliable in comparison with the
diffusion approximation. But the question remains regarding
how rigorous the Eddington approximation is in comparison
with the exact solution of a kinetic equation. 

Using our code and a semi-analytic approach (see Appendix {\ref{appa}) we calculated the critical ratio by solving the
complete system of radiation hydrodynamics equations with no
closure condition. The critical ratio in this case is $(p_r/p_g)_{cr} \simeq 4.5$ we calculated the critical ratio by solving the
complete system of radiation hydrodynamics equations with no
closure condition. The critical ratio in this case is \citet{Belokogne1959}. In contrast
to the Eddington approximation, in our calculation the
Eddington factor is not constant. In the discontinuity region
its value is lower and consequently, lower radiation pressure is
required to reach a continuous solution. The accuracy of our
iterative method is high enough to estimate the critical ratio and
compare it with the previous studies, but further improvements
of the numerical approach are needed both for an exact
calculation of $(p_r/p_g)_{cr}$ and for a consideration of relativistic flows.

{
The diffusion approximation is widely used for the calculation of the critical ratio $(p_r/p_g)_{cr} \simeq 4.45$ by consideration
of the "isothermal jump" \citep{Landau1959, Belokogne1959, Bouquet2000}.
But the "isothermal jump"
is a consequence of the mathematical approximation where the
flow is proportional to the temperature gradient. This
approximation eliminates the temperature jump at the shock.
But the temperature jump occurs due to nonequilibrium
radiation (see \citet{ZeldovichRaizer1966} and Fig. \ref{swedd2}) 
and
the critical ratio becomes higher in more accurate Eddington
approximations ($(p_r/p_g)_{cr} \simeq 8.5$) \citep{ImshennikMorozov1964}.
Our calculations show that both the Eddington approximation
and the M1 closure disregard the contribution of $df/dF$, $df/dP$ terms in the calculation of the critical ratio and overestimate it.
}

In the case where we cannot neglect the energy of the
upstream gas the critical ratio will be lower than its maximum
values (see Fig. \ref{ratio} and a more detailed investigation by \citet{Imshennik1962}). We found a similar situation in test f3. However, test f3 is highly relativistic and requires a separate, more accurate investigation.

\begin{figure}
\includegraphics[width=84mm]{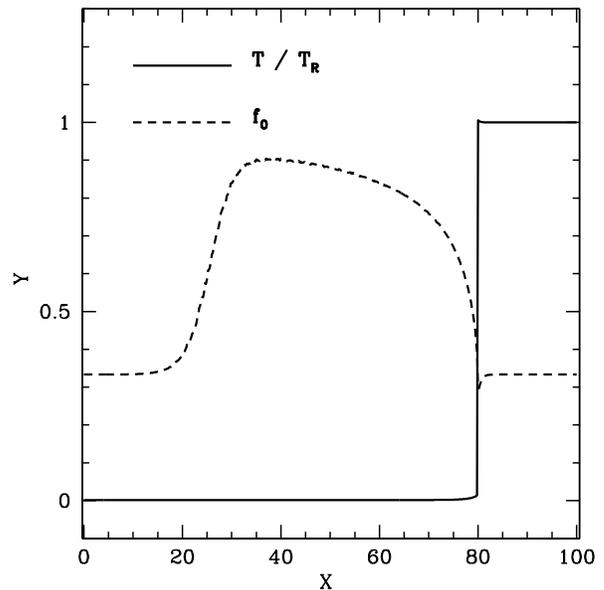}
\caption{Normalized temperature and Eddington factor for test s1 (Table II) in solution with the radiative transfer equation.
 }
\label{t1cmp}
\end{figure}

\begin{figure}
\includegraphics[width=84mm]{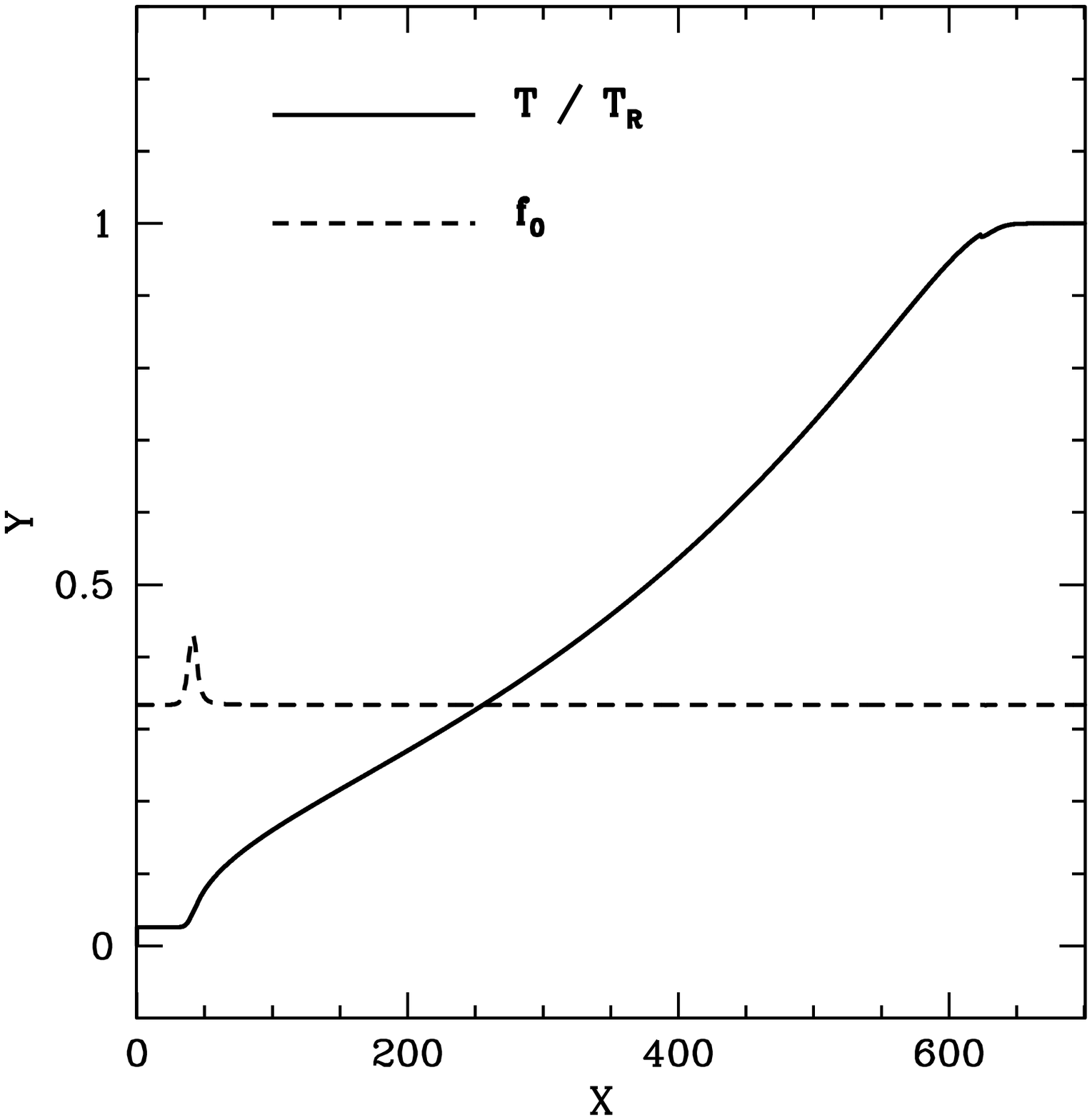}
\caption{Normalized temperature and Eddington factor for test s10 (Table II) in solution with the radiative transfer equation.
 }
\label{t2cmp}
\end{figure}

\begin{figure}
\includegraphics[width=84mm]{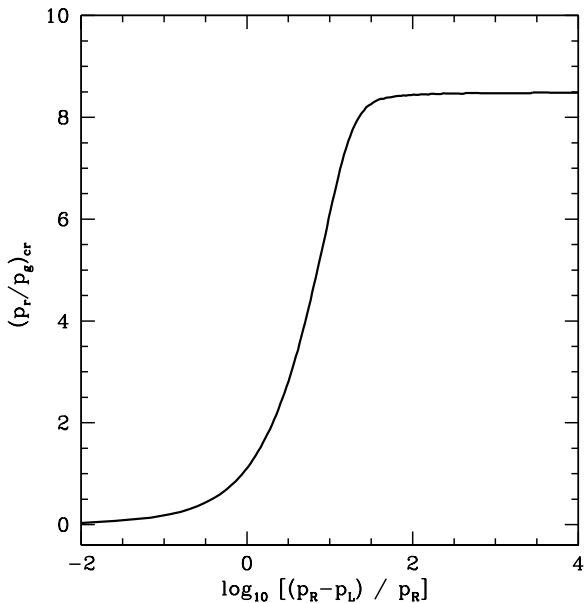}
\caption{Downstream critical radiation-to-gas pressure ratio $(p_r/p_g)_{cr}$ depending on the fractional pressure jump in the Eddington approximation closure.
 }
\label{ratio}
\end{figure}

\section{Conclusions}
\label{sec:conclusions}
\noindent

Our investigation shows that assuming graybody opacity and
the equation of state for an ideal gas in non-relativistic strong
shock waves has a good accuracy in calculating the shock front
structure, regardless of the dominance of radiation. This is not
the case for relatively weak shocks where radiative transfer can
be accurately taken into account to have reliable calculations.
Radiative transfer can affect the gas in the head part of the
shock front and closure approximations cannot already be used
for accurate calculations of radiation fluxes.

The analysis of the criteria for the disappearance of shock
wave discontinuity in radiation-dominated media for strong
shock waves shows that an accurate approach for calculating
the Eddington factors provides the result $(p_r/p_g)_{cr} \simeq 4.5$. This
value is close to the results that were originally derived by  \citet{Belokogne1959}: $(p_r/p_g)_{cr} \simeq 4.45$ for monoatomic gases. Approximations of Eddington factors (e.g., the Eddington, the M1 closure) give the critical ratio $(p_r/p_g)_{cr} \simeq 8.5$ 
\citep{ImshennikMorozov1964}, but for strong shocks radiation and hydro
profiles are not significantly affected. The variations of the
equation of state have a greater influence on the profiles
\citep{Belokogne1972}. 

A complete consideration of the shock wave structure requires
an analysis of relativistic and ultra-relativistic cases along with
scattering in details and will be discussed separately. The
qualitative analysis we have used for shock wave structure can
be extended to relativistic cases, but it requires a higher accuracy
for the solution of the integral-differential equation of radiative
transfer. In our preliminary calculations of relativistic flows for
both the radiative transfer solutions and radiation moment
closure, the critical downstream radiation-to-gas pressure ratio $(p_r/p_g)_{cr}$ becomes larger with increasing velocity of the shock
wave, due to a decreasing of the preheated region. But at highly
relativistic velocities the influence of the Eddington factor
approximation on the shock wave structure becomes more
significant and the usage of a pre-defined radiation moment
closure may lead to incorrect results. 

Our approach in the modeling of the shock wave structure at
arbitrary velocities of the matter is an important step toward
realistic simulations of shock breakouts where the velocity of
matter becomes highly relativistic. This includes simulations of
hypernovae, superluminous supernovae, Type Ibc supernovae,
exploding white dwarfs, GRBs, and their afterglows. 

\section{Acknowledgments}

This research is supported by the World Premier International
Research Center Initiative (WPI Initiative), MEXT,
Japan. The work of S.B. (general formulation of the problem
and numerical algorithms) is supported by the Russian Science Foundation
Grant No. 14-12-00203. KN and AT are supported by the Grant-in-Aid for Scientific Research
of the JSPS (23224004), Japan. 

%This work was partly supported by grant ...
%of the Government of the Russian Federation 11.G34.31.0047 and by grants for
%Scientific Schools 5440.2012.2, 3205.2012.2, and joint RFBR-JSPS grant
%13--02--92119.
%\newpage

%\bibliographystyle{gost705ru}
% \bibliographystyle{gost780ru}
\bibliographystyle{apj}
\bibliography{bibfile}

%%%%%%%%%%%%%%%%%%%%%%%%%%%%%%%%%%%%%%%%%%%%%%%%%%%%%%%%%

\appendix

\section{A. Moment equations of the radiation field}
\label{appa}

For steady flow in plane geometry the radiation field is
described as follows: \citep{Mihalas1980}
\begin{align}
4\pi\frac{d H}{dx} = q^0 = \gamma (q_0^0+\beta q_0^1), \\
\frac{4\pi}{c}\frac{d K}{dx} = q^1 = \gamma (\beta q_0^0+ q_0^1),
\label{4vect}
\end{align}
where $q^{\alpha}$ is a four-vector whose components specify the rate of
momentum and energy exchange between radiation and matter,
\begin{align}
q^0 &= \int_{0}^{\infty}d\nu\oint d\omega [\eta(\mu,\nu)-\chi(\mu,\nu)I(\mu,\nu) ], \\
q^1 &= \int_{0}^{\infty}d\nu\oint d\omega [\eta(\mu,\nu)-\chi(\mu,\nu)I(\mu,\nu) ]\mu, \\
q_0^0 &= 4\pi\int_{0}^{\infty} [\eta_0(\nu_0)-\chi_0(\nu_0)J_0(\nu) ]d\nu_0, \\
q_0^1 &= -4\pi\int_{0}^{\infty} \chi_0(\nu_0)H_0(\nu) ]d\nu_0,
\end{align}
where the subscript $_0$ relates to the comoving frame. 

We assume a graybody form for all opacities 
\begin{align}
\chi_0(\nu_0) = \kappa\rho_0 ,
\end{align}
\begin{align}
\kappa = \kappa_{abs} + \kappa_{sc} = \kappa_{abs} ,
\end{align}
where
$\kappa$ is frequency- independent opacity and $\rho_0$ is the rest mass energy
density.  Scattering opacity is understood to be
zero.
The thermal emissivity $\eta_0(\nu_0)$ and the absorption coefficient $\chi_0(\nu_0)$ are related by
Kirchhoff's law $\eta_0 = \chi_0 B_0$:
\begin{align}
 q_0^0 &= 4\pi\chi_0 (B_0-J_0) = \kappa\rho_0(4\pi B_0-E_0), \\
 q_0^1 &= -4\pi\chi_0 H_0 = -\kappa\rho_0 F_0,
\end{align}
where $B$ is the intensity the in thermal equilibrium.

Substituting components of the four-vector in (\ref{4vect}),
moment equations can be written in the following form:
\begin{align}
&  \frac{dF}{dx}=\frac{\kappa U_1}{\beta}\Big( 
{a_R} T^4+\frac{\beta F(2-f-\beta^2)-P(1-f\beta^2)}{f-\beta^2}
 \Big), \\
% \label{r1}
&  \frac{dP}{dx}=\frac{\kappa U_1}{\beta}\Big( 
({a_R} T^4+P)\beta-F
 \Big),
 \label{r2} 
\end{align}
{where $a_R$ is the radiation constant }and for temperature $T$ we have the following expression:
\begin{align}
  T = {\frac{m}{k_B}}\frac{p}{\rho} =
  {\frac{m}{k_B}}\Big( \frac{U_3-\gamma U_1 - F}{U_1\gamma(n+1)} \Big),
\end{align}
{where $m$ is the mean mass of baryons in the fluid, $k_B$
is Boltzmann's constant. Hereafter we set $k_B=1$.}

The elimination of $p$ and $\rho$ from the radiation hydrodynamics equations gives the following equation for $\beta$:
\begin{align}
  H(\beta,F,P,n) = -U_1/\gamma-(n+1)\beta(U_2-P)+(1+n\beta^2)(U_3-F) = 0.
\end{align}

Let us perform phase analysis in the $PF$ plane.
After some algebra equilibrium points $F_x$ and $P_x$ 
of the system (1)-(2) can be found from the 
following algebraic equation
\begin{align}
  H(\beta_x,0,0,n)(1-nf)^3 = \beta_x a_R m^4 \Big[\frac{f\gamma_x H(\beta_x,0,0,1/f)}{U_1}\Big]^4.
\label{veqn} 
\end{align}

Finding $\beta_x$ from this equation, $F_x$ and $P_x$ can be determined from:
\begin{eqnarray}
 &&  F_x=\frac{(1+f)\gamma_x^2 H(\beta_x,0,0,n)}{1-nf},
\end{eqnarray}   
\begin{eqnarray}
 &&   P_x \beta_x= F_x-\beta_x a_R(T(\beta_x,F_x,n))^4.
\end{eqnarray}   
   
The equation (\ref{veqn}) has several roots, but only two of them have
physical meaning with non-negative pressure. These roots can be used for the
construction of shock tube configuration.

Let us write the equations (\ref{r2}) in linear form in 
the neighbourhood of an equilibrium point:
\begin{align}
  \frac{dF}{dx}=a(P,F)(P-P_x)+b(P,F)(F-F_x),
 \label{lin2}
\end{align}
\begin{align}
  \frac{dP}{dx}=c(P,F)(P-P_x)+d(P,F)(F-F_x),
 \label{lin}
\end{align}
 where coefficients are equaled to
\begin{align}
  a = \kappa U_1 \Big[ \frac{a_R T^3}{\beta^2}\Big(4\beta
 \frac{\partial T}{\partial P} &-T\frac{\partial \beta}{\partial P}\Big) +\frac{f\beta^2-1}{\beta(f-\beta^2)}+ \nonumber \\
&+ \Big(\frac{4\beta^3(1-f)F + 
  [f+(f^2-3)\beta^2+f\beta^4]P}{\beta^2(f-\beta^2)^2}
 \Big)\frac{\partial \beta}{\partial P} 
  +\Big( \frac{(1+\beta^2)P-2\beta F}{ \gamma^2\beta (f-\beta^2)^2}   \Big)
 \frac{\partial f}{\partial P} \Big],
\end{align}
\begin{align}
 b = \kappa U_1 \Big[ \frac{a_R T^3}{\beta^2}\Big(4\beta
 \frac{\partial T}{\partial F} &-T\frac{\partial \beta}{\partial F}\Big) +
 \frac{2-f-\beta^2}{(f-\beta^2)}+ \nonumber \\
 &+\Big(\frac{4\beta^3(1-f)F + 
  [f+(f^2-3)\beta^2+f\beta^4]P}{\beta^2(f-\beta^2)^2}
 \Big)\frac{\partial \beta}{\partial F} 
  +\Big( \frac{(1+\beta^2)P-2\beta F}{ \gamma^2\beta (f-\beta^2)^2}   \Big)
 \frac{\partial f}{\partial F} \Big],
\end{align}
\begin{eqnarray}
 && c = \kappa U_1 \Big[ 4aT^3\frac{\partial T}{\partial P} +1+
 \frac{F}{\beta^2} \frac{\partial \beta}{\partial P} \Big],
\end{eqnarray}
\begin{eqnarray}
 && d = \kappa U_1 \Big[ 4aT^3\frac{\partial T}{\partial F} -\frac{1}{\beta}+
 \frac{F}{\beta^2} \frac{\partial \beta}{\partial F} \Big].
\end{eqnarray}    

From algebraic equations (3-4) we can find:
\begin{eqnarray}
 && \frac{\partial\beta}{\partial P} = - \frac{(n+1)\beta}
     {G(\beta,F,P,n)}, 
\end{eqnarray}
\begin{eqnarray}
 && \frac{\partial\beta}{\partial F} = \frac{1+n\beta^2}
     {G(\beta,F,P,n)}, 
\end{eqnarray}
\begin{eqnarray}
 && \frac{1}{m}\frac{\partial T}{\partial P} = \gamma\beta^2
 \frac{U_3-F}{U_1 G(\beta,F,P,n)}, 
\end{eqnarray}
\begin{eqnarray}
 && \frac{1}{m}\frac{\partial T}{\partial F} = 
     - \frac{G(\beta,F,P,n)+\beta\gamma^2(1+n\beta^2)(U_3-F)
      }{(n+1)\gamma U_1 G(\beta,F,P,n)}, 
\end{eqnarray}
\begin{align}     
  G(\beta,F,P,n)
  = \beta\gamma U_1+2n\beta(U_3-F)-(n+1)(U_2-P).
\end{align}

The sign of the determinant $P=ad-bc$ determines the nature of the
singular points $F_x$,$P_x$. If $P<0$, we have a saddle point, $P>0$ --- node ($(a+d)^2 > P$) or focus ($(a+d)^2 < P$).

Sadle singular points can not be connected by one
separatrix \citep{Andronov1987} and the boundary value problem must be 
solved in this case to find the hydro and radiation quantity profiles.

For the M1 closure the analysis is similar to the Eddington
closure case because the derivatives $\partial f / \partial F$,$\partial f / \partial P$ 
for the M1
closure are proportional to radiation flux in the comoving frame
$F_0$ and in the equilibrium points equal to zero

In the solution of the radiative transfer equation the
Eddington factor $f$ is not constant and the singular points
behavior is different from the Eddington closure case ($f=1/3$). 

Partial derivatives $\partial f/\partial F$ and $\partial f/\partial P$ can be 
found from 
\begin{align}
df = \frac{\partial f}{\partial F} dF + \frac{\partial f}{\partial P} dP 
\end{align}
and coupled with 
\begin{align}
\frac{dF}{dP} = \frac{c}{d-l},
\end{align}  
where $l$ is an eigenvalue of the matrix of the linearized system (\ref{lin2}-\ref{lin}).

The total derivatives $df/dF$ and $df/dP$ can be estimated from radiative transfer
equation in the limit of small velocities $\beta$. In the case of a gas-dominated flow the solution
of the radiative transfer equation can be written as a formal solution (see Appendix \ref{appb}),
and for an asymptotic solution, where $dE_0/dx=|dF_0/dx|=dP_0/dx$ we have
\begin{align}
  \frac{df}{dP} = \frac{d(P_0/E_0)}{dP} = \frac{2}{3E_0} \frac{dP_0}{dP},   \\
  \frac{df}{dF} = \frac{d(P_0/E_0)}{dF} = \frac{2}{3E_0} \frac{dP_0}{dF},  
\end{align}
Finally, $dF$ and $dP$ can be found by differentiating the Lorentz transformations (\ref{moments_transform}) and the equation for velocity:
\begin{align}
    dF = \gamma^4(4\beta F_0+(1+\beta^2)(P_0+E_0))d\beta
    + \gamma^2((1+\beta^2)dF_0+\beta(dP_0+dE_0))
\end{align}
\begin{align}
    dP = 4\gamma^4(\beta(E_0 + P_0)+(1+\beta^2)F_0)d\beta
    + \gamma^2(2\beta dF_0+dP_0+\beta^2 dE_0))
\end{align}
\begin{align}
d\beta = \frac{\partial \beta}{\partial F} dF + \frac{\partial \beta}{\partial P} dP 
\end{align}

For the radiation dominated flow $dE_0/dP_0$ and $dF_0/dP_0$ can be found by a numerical solution of the radiative transfer equation.

\section{B. Formal solution of the equation of radiative transfer for a step function source function}
\label{appb}

The formal solution of the equation of transfer in a static
medium of the layer with a finite optical thickness (see \citet{Chandrasekhar1950}) reduce to
\begin{align}
 I(\tau,+\mu,\phi) = I(\tau_1,\mu,\phi) \exp^{-(\tau_1-\tau)/\mu} 
 +\int_{\tau}^{\tau_1} 
 S (t,\mu,\phi) \exp{-(t-\tau)/\mu} \frac{dt}{\mu},
\end{align}
\begin{align}
 I(\tau,-\mu,\phi) = I(0,-\mu,\phi) \exp^{-\tau/\mu} 
 +\int_{0}^{\tau} 
S (t,-\mu,\phi) \exp{-(\tau-t)/\mu} \frac{dt}{\mu}.
\end{align}
where $S(t,\mu,\phi),(1\ge\mu\ge 0)$ is the source function. 
If we assume Kirchhoff's law and a step source function 
\begin{align}
     S = B_R, (x>0), \\
     S = B_L, (x<0),
\end{align}
then intensities and radiation moments can be found as:
\begin{align}
  I_{R,-} = B_R,
\end{align}
\begin{align}
I_{R,+} &= B_L \exp {(-\tau_R / \mu) }+ 
B_R \int_0^{\tau_R} \exp{(-(\tau_R-t)/\mu) }\frac{dt}{\mu},
\end{align}
\begin{align}
 J_{R} &= \frac{1}{2}\int_{-1}^{1} I_R d\mu =
\frac{1}{2}\int_{-1}^{0} I_{R,-} d\mu + \frac{1}{2}\int_{0}^{1} I_{R,+} d\mu = \nonumber \\ 
 &= \frac{1}{2}\Big[ B_R + B_L \int_{0}^{1} \exp{(-\tau_R/\mu)} d\mu 
                    + B_R \int_0^{\tau_R} \exp{(-(\tau_R-t)/\mu) }\frac{dt}{\mu}  \Big]                      
% = \frac{1}{2}\Big[ B_R + B_L E_2 (\tau_R) + B_R \int_{0}^{\tau_R}E_1(t)dt  \Big] 
 = \frac{1}{2}\Big[ 2B_R + (B_L - B_R) E_2 (\tau_R)   \Big], \\
 H_{R} &= \frac{1}{2}\int_{-1}^{1} I_R \mu d\mu =
\frac{1}{2}\int_{-1}^{0} I_{R,-} \mu d\mu 
+
\frac{1}{2}\int_{0}^{1} I_{R,+} \mu d\mu =
%&& = \frac{1}{2}\Big[ B_R + B_L \int_{0}^{1} \exp{(-\tau_R/\mu)} d\mu + \nonumber \\
%&&                    + B_R \int_0^{\tau_R} \exp{(-(\tau_R-t)/\mu) }\frac{dt}{\mu}  \Big] = 
%                     \nonumber \\ 
% = \frac{1}{2}\Big[ -\frac{B_R}{2} + B_L E_3 (\tau_R) + B_R \int_{0}^{\tau_R}E_2(t)dt  \Big] 
  \frac{1}{2}\Big[ (B_L -B_R) E_3 (\tau_R)   \Big], \\
 K_{R} &= \frac{1}{2}\int_{-1}^{1} I_R \mu^2 d\mu =
\frac{1}{2}\int_{-1}^{0} I_{R,-} \mu^2 d\mu  
+\frac{1}{2}\int_{0}^{1} I_{R,+} \mu^2 d\mu =
  \frac{1}{2}\Big[ \frac{2}{3} B_R + (B_L - B_R) E_4 (\tau_R)   \Big].
\end{align}

The values of the radiation moments at $x<0$ can be found similarly.

\end{document}